\documentstyle[aps,multicol,epsfig]{revtex}

\begin{document}

\draft

\title{Universal behavior of relaxational heterogeneity in
glasses and liquids}

\author{D. Caprion, J. Matsui and H.R. Schober}

\address{Institut f\"ur Festk\"orperforschung, Forschungszentrum J\"ulich, 
D-52425 J\"ulich, Germany}

\date{\today}

\maketitle

\begin{abstract}
We report an investigation of the heterogeneity in super-cooled liquids and
glasses using the non-Gaussianity parameter. We simulate 
selenium and a binary Lennard-Jones system
by molecular dynamics. In the non-Gaussianity
three time domains can be distinguished. First there is an increase
on the ps-scale due to the vibrational (ballistic) motion of the atoms.
This is followed, on an intermediate time-scale, by a growth,
due to local relaxations ($\beta$-relaxation) at not too high 
temperatures. 
A maximum is reached at times corresponding to long range diffusion
($\alpha$-relaxation). At long times the non-Gaussianity slowly drops,
the system becoming homogeneous on these time-scales.
In both systems studied,
the non-Gaussianity follows in the intermediate
time domain, corresponding to the $\beta$-relaxations, a law 
$\propto \sqrt{t}$.
This general behavior is explained by collective hopping and dynamic
heterogeneity. We support this finding by a model calculation.
\end{abstract}


\begin{multicols}{2}

Although glass is one of the most common materials, its physics and 
especially its dynamics are still only poorly understood. In addition to
sound waves, two level systems \cite{phillips:72,anderson:72} and
quasi-local (resonant) vibrations \cite{LS:91}, experiments indicate a
wide
distribution of relaxations, i.e. non-periodic changes of the local
structure \cite{hunklinger:86,buchenau:88}. In super-cooled liquids one
observes, apart from the vibrational (ballistic) motion of the atoms,
two types of relaxations with different time-scales.
These $\beta$- and $\alpha$-relaxations are attributed to short range
(cage) motion and diffusion, respectively.

In recent years these relaxations, both in glasses and in liquids, 
have been studied intensively by
experiment \cite{arbe:98,wuttke:96,meyer:98,zorn:97} 
and theoretically
\cite{miyagawa:88,kob:95b,kob:97,doliwa:98}. 
One particular aim was to determine
whether the
relaxations involve only groups of atoms or are spread 
over the whole
system. The first case, where
relaxations are restricted to a few atoms only,
is known as heterogeneous scenario; the other one 
as homogeneous.

Spatial heterogeneity is thought to be responsible for
the non-exponential relaxations in super-cooled liquids \cite{hetero}.
This view has recently been challenged on the basis of inelastic neutron
scattering experiments on polymers \cite{arbe:98}, however, see also 
\cite{heuer:99c}.

To understand its effects, it is necessary to know the properties of the
``dynamic heterogeneity'' itself, e.g. the time and temperature dependence. 
Qualitatively it is known \cite{arbe:98,kob:97,CS:00}, 
that the system becomes homogeneous at all 
temperatures
for sufficiently long times, corresponding
to the $\alpha$-regime. 
In the intermediate time domain, corresponding to the
$\beta$-relaxation, heterogeneity becomes more pronounced
when the system is cooled down. Here we will show that there is a
universal law governing heterogeneity at these time-scales.

This is closely related to collectivity of motion.
Measurements of the isotope effect have shown that diffusion both in 
glasses and in super-cooled liquids is highly collective 
\cite{faupel:90,ehmler:98}. 
A similar very small isotope effect was also observed in 
simulations of a Lennard-Jones liquid \cite{KS:00}. In glasses, one
observes collective jumps of chain-like structures \cite{SOL:93,OS:99,OS:95}.
Similar mobile structures are also observed in the under-cooled liquid
\cite{SGO:97,donati:98}.

Following our previous work \cite{CS:00}, we investigate the
non-Gaussianity behavior, i.e. the heterogeneity, of relaxations in
glasses.
In this letter we focus especially on intermediate times,
shorter than the typical diffusion time.
As shown previously,
the non-Gaussianity increases markedly in this time domain.
Here we want to go
one step further.
First we show that molecular dynamics simulations of two different
systems,
Se and binary Lennard-Jones (LJ), give strikingly the same law
for the increase of the non-Gaussianity parameter (NGP), $\alpha_2(t)$
defined further down. 
In this intermediate
time domain the non-Gaussianity 
follows a power law $\alpha_2(t) \propto \sqrt{t}$, for both systems 
and both for temperatures 
above and below $T_g$. 
We propose a simple model based on 
the collective hopping of groups of particles.

The simulations for Se and LJ were both done
with a velocity-Verlet algorithm, controlling the temperature by
velocity adjustment and using the equilibrium volume at the given  temperature,
i. e. zero average pressure. 

We describe Se with a 3-body potential \cite{OJRS:96}. 
This potential has been used
previously to calculate vibrations\cite{OS:93}, local
relaxations \cite{OS:95,SGO:97} and heat transport \cite{olig:99}
in amorphous Se. It  provides
a sound basis for the study of both the atomistic and the electronic
structure \cite{koslowski:99}. We prepared 4 independent samples of
hot liquid, each containing $2000$ 
atoms. These were then quenched to the desired
temperatures
with rates of $10^{13}$~K/s. Before using the
configurations for the measurements, they were aged for several ns.
The effective quench rates were thus of order $10^{10}$~K/s. The glass
transition temperature is estimated as $T_{\rm g} \approx 300$~K, and the  
critical temperature $T_c \approx 330$~K.
More details 
are given in Ref.~\onlinecite{CS:00}.

For the binary LJ simulations we take the frequently used parameters 
of Kob and Anderson \cite{kob:97,kob:95a}.
The simulations are done with 5488 atoms and a composition of 20\% 
small particles. 
We quench from appropriately aged samples with a rate of
about $10^{11}$~K/s (relating the LJ-values to Ar)
and subsequently age the samples again. 
A heat bath was simulated by comparing the temperature averaged over 20
time steps with the nominal temperature. At each step 1\% of the temperature
difference was adjusted by random additions to the particle velocities.
To ease comparison
with previous LJ simulations we shall, in the following,
use the usual LJ units.
Details of the simulation procedure are analogous to the ones described for 
the monatomic LJ system \cite{KS:00}. 
From the diffusion constant we determine $T_c \approx 0.37$.
The discrepancy of this value compared to the one in previous simulations
\cite{kob:95b,kob:95a}, is due to the difference of densities. The
previous work was done for a constant density of $\rho = 1.2$ whereas
we find for our equilibrium samples at $T=T_c$ a lower value of
$\rho = 1.15$. We also used a slightly larger cut-off of the potentials.

\begin{figure}
\label{fig1}
\epsfig{figure=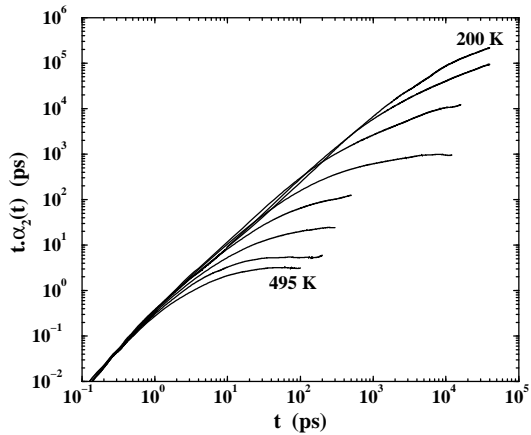,width=7cm}
\caption{Non-Gaussianity parameter multiplied by time against time
in a log-log representation. The values are obtained from a molecular
dynamics simulation of Se at the temperatures 
(from bottom to top): 495~K, 445~K, 400~K, 355~K, 330~K, 290~K, 255~K
and 200~K.}
\end{figure}

To quantify the heterogeneity of the
relaxations we follow previous work
\cite{zorn:97,miyagawa:88,kob:95b,doliwa:98,odagaki:99}
and use the 
non-Gaussianity parameter (NGP)\cite{rahman:64}:
\begin{equation}
\label{ngp}
\alpha_2(t)=\frac{3 <\Delta r^4(t)>}{5 <\Delta r^2(t)>^2}-1,
\end{equation}
where $<...>$ denotes time averaging, $\Delta r^2(t)$ is the mean
square
displacement and $\Delta r^4(t)$ is the mean quartic displacement.
Experimentally the NGP can be determined from the $q$-dependence
of the Debye-Waller factor \cite{zorn:97}. From simulations 
the qualitative behavior of $\alpha_2(t)$ is well known. 
Starting from $\alpha_2(t=0) = 0$, it rises 
on a time-scale typical for vibrations ($t \approx 1$~ps) 
to values around $\alpha_2 = 0.2$. In a hot liquid the NGP drops from
this value on a ps-scale. In under-cooled liquids and in
glasses the NGP keeps growing on intermediate time-scales and
reaches values an order of magnitude larger. Only on the time-scale
of diffusion or $\alpha$-relaxation does the NGP drop and finally
reaches, for $t \to \infty$, the limit  $\alpha_2 = 0$. 
This latter limit reflects the ergodicity of the system for long times.
From the increase, one concludes that
the relaxations are mainly heterogeneous in the
intermediate time scale. This becomes more and
more pronounced as the system is cooled down \cite{CS:00}. 

Here we want to stress a material independent
property in this time  domain. For this,
we plot for Se $\alpha_2(t)$ multiplied by $t$ against time in a log-log
representation, Fig.~1. The most interesting feature of
this plot is the appearance of a master curve for all temperatures
stretching over a time domain from 
$10^{-1}$~ps to $10^{3}$~ps, i.e. four orders of magnitude. 
This master
curve is independent of any rescaling constants or procedures.
It corresponds to a power law $t^{3/2}$ leading to
$\alpha_2(t) \propto \sqrt{t}$.

To check whether this behavior results from the particular structure of 
amorphous Se, which is constituted from chains and rings, we repeated
the calculations for a binary LJ system. This model is frequently used
as an idealized dense packed metallic glass. The nearest neighbor coordination
is near 12 rather than 2 in Se. 
We plot the NGP
of binary
LJ in the same way as for Se, i.e. $t\cdot\alpha_2(t)$ versus $t$ in
a log-log plot. As Fig.~2 shows, the time dependence  of the NGP 
follows the same power law $\alpha_2(t) \propto \sqrt{t}$ as
seen in Se. Moreover, we observe this behavior not only for the average
NGP, shown in Fig.~2, but also for both components separately. 

\begin{figure}
\label{fig2a}
\centerline{\hbox{\epsfig{figure=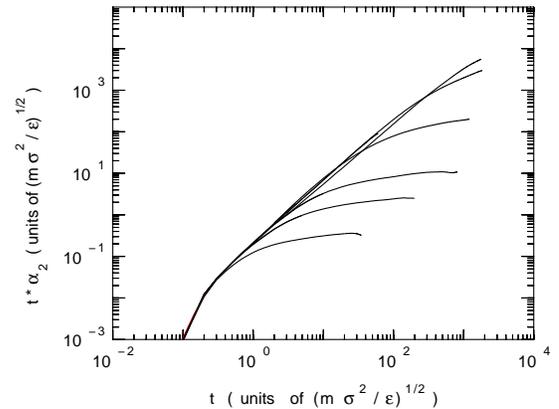,width=8cm}}}
\caption{Non-Gaussianity parameter multiplied by time versus time
obtained
from molecular dynamics simulation of a binary Lennard-Jones system at the
temperatures (from bottom to top): 0.88, 0.56, 0.48, 0.40, 0.36 and 0.32.}
\end{figure}

From the above we conclude
that the NGPs of different types of structural
glasses and super-cooled liquids follow at intermediate times the 
same time dependence: $\alpha_2(t) \propto \sqrt{t}$.
Therefore, we think that the mechanism responsible for the
increase
of the NGP, i.e. of the heterogeneity, is
common to all kinds of glass-forming materials.

In our previous investigation of the non-Gaussianity \cite{CS:00}, we
have clearly shown that the increase of non-Gaussianity is due to
relaxations. Moreover it has also been shown that in under-cooled
liquids
and in the glassy phase, clusters of so called mobile
particles exist \cite{OS:99,OS:95,SGO:97,donati:98}.
Theses move in a given time over greater distances than the 
average. On the other hand we also
know from experiments \cite{arbe:98} and simulations \cite{CS:00}
that
on the long time scales of diffusion the NGP drops, which indicates
that the heterogeneity decays. 

Starting from these two points,
we build a simple model to explain the time dependence of the NGP. 
We make three assumptions, all based on previously known results.
First, all atoms have a vibrational mean square displacement, increasing 
with temperature $\propto T$ and giving an initial $\alpha_2(t) \approx 0.2$.
Secondly, there are groups of  mobile atoms which jump collectively. 
And thirdly, after such a jump
some atoms can leave while others enter a mobile group.
Fig.~3 depicts such a collective jump schematically.

\begin{figure}
\label{fig3a}
\centerline{\hbox{\epsfig{figure=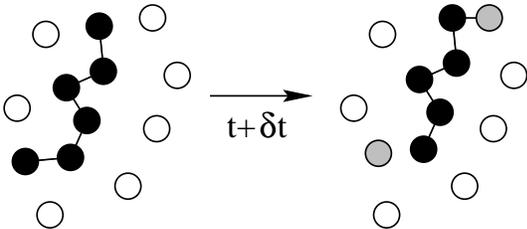,width=7cm}}}
\caption{Schematic representation of a collective jump (left configuration
to right configuration).
Mobile and 
immobile atoms are indicated by full and empty circles, respectively.
The grey circles show the atoms which have changed their group.
After the jump one previously mobile atom has left the group,
i.e. it will not participate in the next jump. Another atom has joined
the group instead.} 
\end{figure} 

In order 
to make this model more tractable we use the most simple approximations.
We have, however, checked carefully that the results do not depend on
these details. We consider a system formed of several groups,
each group containing 10 atoms.
Each of these groups
of atoms can jump collectively over some barrier into an adjacent
configurational minimum position of the underlying energy landscape 
\cite{heuer:97b,sastry:98,schulz:98,bhatta:99}. 
For simplicity we take a constant 
distribution of  the activation energies.
Furthermore, we take a constant probability for jump reversal, ranging from
0.1 to 0.5.
A backward jump means that the group of atoms returns
to the previous positions. 
Each time a group of atoms jumps, one atom
of the group is exchanged with an atom of another randomly chosen group.
This accounts approximately for the changed
local environment of the atoms after the jump
which will cause some atoms to be in more stable neighborhoods and some
others to become more unstable in turn.
In the long time limit, this exchange
procedure 
leads to the reduction of the
heterogeneity since each atom will ultimately participate in the diffusion. 
To be effective, an energy barrier has to have a minimum hight, depending on
temperature.
If the system is at a sufficiently high temperature the hopping
over the lowest barriers will
merge with the anharmonic vibrations. The atoms will instead be restrained
by the next barrier of sufficient hight. 
Seen the other way round, when the system is cooled down, it 
becomes affected by more and
more fine details of the energy landscape. 
To take this into account, we 
define at each temperature a  minimum hight for a barrier to be effective for 
relaxations, $E_{\rm min} = k_{B}T$. Lower barriers are replaced randomly
by barriers above this thresh-hold, keeping a flat distribution of barrier
hights.
Finally the vibrations are accounted for by an ``instantaneous'' mean
square displacement of the atoms. 
With this simple model in mind, we perform Monte Carlo-like simulations.

\begin{figure}
\label{fig3}
\centerline{\hbox{\epsfig{figure=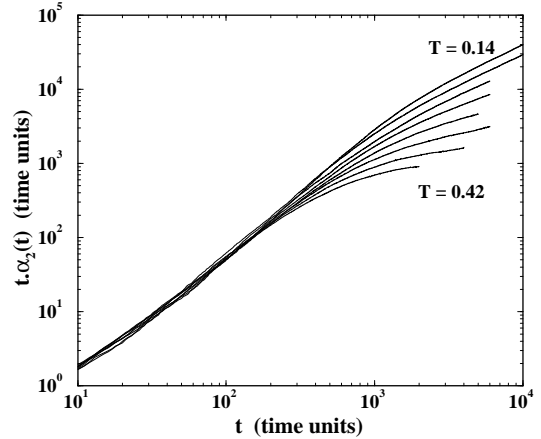,width=7cm}}}

\caption{Non-Gaussianity parameter multiplied by time versus time,
computed
for a simple model of relaxation (see text) at different temperatures
in the super-cooled liquid and in the glass.}
\end{figure}

We build a system of 2000 atoms, divided into 200 groups having different
activation energies. At each time step 
we check for each group $i$ whether
this
group will jump with the associate probability $\exp{(-E_i/k_BT)}$.
If the group jumps, we move all the atoms of this group, 
and then exchange one atom of
the group 
with another atom. We have checked that neither the exchange rate nor
the probability of jump reversal
change the
exponent of the power law. They merely affect the absolute value of the
non-Gaussianity
parameter. We repeat this procedure for several time steps and starting
configurations. During these simulations the non-Gaussianity
parameter is computed
at different temperatures from 0.42, which in our units
corresponds to
the liquid, to 0.14 which is under an equivalent $T_g$. Fig.~4 shows
the result of these simulations.

As one can see this simple model reproduces the power law
$\alpha_2(t) \propto \sqrt{t}$, found in the molecular dynamics 
simulation, which is striking due to the
simplicity of the model and even more so to  the fact
that all details of the interaction in the
materials are neglected.
In other words, the model can be applied to any kind of glass-forming material,
once one
assumes that the relaxations are by collective hopping of groups of particles.
Therefore, we think that the behavior of the non-Gaussianity parameter is
universal and will also be observed in oxide glasses such as silica and in
polymeric glasses.

In this work
we have concentrated on the intermediate time regime which
is more easily accessible to experiment. There remain some other
interesting questions. Comparing Fig.~2 with the predictions of the
trapping diffusion
model \cite{odagaki:94}
we see a difference for long times. The
decay of the NGP in the simulation is clearly slower than the
predicted $1/t$.
The reason for this is not yet understood. Another open question
is the NGP at temperatures much below $T_g$. Lowering the temperature
the spectrum of the activation energies should eventually
become important. 

To conclude we have presented the results of two independent molecular
dynamics simulations on completely different systems
and
of a simple model. All these results show the same 
power
law for the non-Gaussianity in the intermediate time range, corresponding
to
the $\beta$-relaxations in under-cooled liquids and in glasses. 
This increase of the non-Gaussianity, i.e. of
the heterogeneity of the relaxations, proportional to $\sqrt{t}$,
can be understood as resulting from the collective hopping of  
groups of particles.
Assuming that this mechanism is common to all kinds of
glass-formers, we believe that heterogeneity will always increase in the
intermediate time
regime domain following the power law $\sqrt{t}$, at all temperatures
and in
all kinds of materials.

We would like to thank C. Oligschleger for her help at the beginning of
the
simulation on selenium. For financial support we thank
the Alexander von Humboldt foundation (DC) and Kyushu University (Japan) 
(JM).

\end{multicols}

\end{document}